\documentclass{article}
\usepackage{spconf,amsmath,graphicx,hyperref}
\usepackage{color}
\usepackage{booktabs}
\usepackage{amsfonts}

\ninept

\title{Exploring Representation Autoencoders for Text-to-Audio Generation}

\name{Bernardo V. Miranda \quad Hugo Malard \quad Gaël Richard \quad Mathieu Fontaine \quad Slim Essid\sthanks{This work is independent from S. Essid’s role at NVIDIA.}}
\address{LTCI, Télécom Paris, Institut Polytechnique de Paris, France}

\begin{document}

\maketitle

% Abstrack
% ---------------------------------
\begin{abstract}
Representation autoencoders (RAE), created by pairing a frozen pretrained self-supervised learning (SSL) encoder with a trained decoder, have recently demonstrated state-of-the-art performance on image generation. Their use in the audio domain, however, remains limited: text-to-audio (TTA) models are typically trained in the latent space of dedicated autoencoders where the encoder and the decoder are trained jointly with reconstruction objectives. This paper presents a study on audio representation autoencoders (AuRA) and their applicability to general audio TTA. We train decoders for two pretrained SSL audio encoders and evaluate their reconstruction quality against autoencoders used by recent TTA systems. After obtaining decoders for the SSL representations, we also train rectified flow models in the SSL and baseline latent spaces to isolate the autoencoder's effect on generation. We find that generation in SSL spaces achieves comparable or superior performance to generation in the baseline spaces, despite differences in reconstruction performance. These results highlight that reconstruction quality alone is not a sufficient proxy for generative performance.
\end{abstract}

% Keywords
% ---------------------------------
\begin{keywords}
text-to-audio generation, representation autoencoders, self-supervised learning
\end{keywords}

% Sections
% ---------------------------------
\section{Introduction}
\label{sec:introduction}

Since their introduction in~\cite{ldm}, latent diffusion models (LDM) have become the standard approach for image and audio generation. With LDMs, the generation paradigm shifted from modeling data distributions directly in sample space to modeling them in the compact, learned, latent space of an autoencoder. Text-to-audio (TTA) generation was no exception to this trend, and current state-of-the-art TTA models~\cite{stable-audio, stable-audio-open, stable-audio-2, stable-audio-3, tango-flux, audioldm2} generate new samples in the latent spaces of dedicated autoencoders trained with reconstruction losses.

LDMs rest on the assumption that generation is easier in a compact space: by discarding perceptually irrelevant detail, the autoencoder yields a lower-dimensional distribution that a generative model can capture more efficiently~\cite{ldm}. This assumption, however, has been challenged in the image domain by generative models operating --- and achieving competitive performance --- on spaces of far higher dimensionality than typical LDM latents~\cite{jit, rae}, suggesting that less compact spaces could be equally suited for generation. 

Representation autoencoders (RAE)~\cite{rae} are built on the hypothesis that self-supervised learning (SSL) representation spaces are well adapted for generative modeling. In the image domain, RAEs are created by pairing a pretrained, frozen, SSL encoder with a trained decoder that maps SSL representations back to pixel space. Intuitively, as SSL representations are commonly used in computer vision for understanding tasks~\cite{dinov2}, they carry rich semantic information, which makes them plausible candidates for generation despite their dimensionality. With RAEs, new samples are generated directly in the SSL representation space, which replaces the usual compressed latent space of LDMs. 

RAEs achieve state-of-the-art performance in image generation, but are still relatively unexplored in the audio domain. Although RAEs have been applied to symbolic music~\cite{midi-rae-jepa}, to the best of our knowledge, no TTA system today operates on SSL representation spaces directly. In~\cite{stable-audio, stable-audio-open, stable-audio-2, tango-flux} generation is done in the latent space of a convolutional variational autoencoder (VAE) trained with reconstruction and adversarial losses. Inspired by RAEs, the transformer autoencoder of~\cite{stable-audio-3, same} uses higher dimensional latents and auxiliary generative, semantic regression, and contrastive losses to obtain a ``semantically aligned'' generation space. Nonetheless, it employs strong time downsampling, reconstruction losses, and jointly trains the encoder and decoder. Finally, in~\cite{audioldm2} generation conditioning is given by features similar to AudioMAE's~\cite{audio-mae}; however, new samples are also created in the latent space of a convolutional VAE.

\begin{figure}[htb]
\centering
\includegraphics[width=0.85\columnwidth]{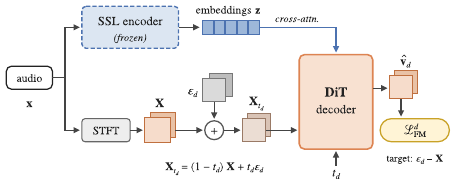}
\caption{Audio reconstruction using a DiT conditioned on the SSL embeddings.}
\label{fig:reconstruction}
\end{figure}

\begin{figure*}[t]
\centering
\includegraphics[width=0.85\textwidth]{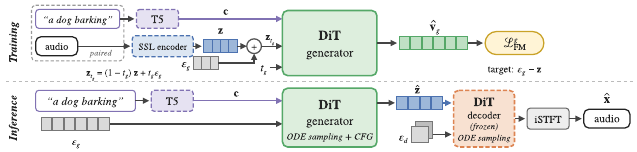}
\caption{Text-to-audio generation using AuRAs. A generator is trained to create new SSL embeddings conditioned on text; at inference time, new embeddings are sampled with the generator and then decoded with the DiT described in Section~\ref{sec:aura-dec}. }
\vspace{-0.2cm}
\label{fig:generation}
\end{figure*}

In this context, this paper provides a study on audio representation autoencoders (AuRA) and their applicability to general TTA. First, we pair two frozen SSL audio encoders --- MATPAC++~\cite{matpac++} and BEATs~\cite{beats} --- with decoders trained to recover original audio signals from their representations in these two spaces (Fig.~\ref{fig:reconstruction}). The two AuRA pairs are then compared, using perceptual and signal-based metrics, to the autoencoders of Stable Audio Open~\cite{stable-audio-open}, Stable Audio 3~\cite{stable-audio-3, same}, and AudioLDM2~\cite{audioldm2} in terms of reconstruction quality. Second, to isolate the autoencoder's impact on TTA, we train text-conditioned rectified flow models on the two AuRAs (Fig.~\ref{fig:generation}) and three baseline latent spaces. TTA outputs are evaluated objectively for caption adherence and overall audio quality. We additionally assess model robustness to outlier captions and scaling properties of AuRAs.

Results show that AuRAs achieve equivalent or superior overall audio quality and caption adherence, despite inferior reconstruction performance. Furthermore, they are comparable to the baselines in terms of robustness and obtain slightly faster convergence per training sample, but require more compute in each forward pass due to longer token sequences in SSL representations. Code, pretrained weights, and audio examples are available online.\footnote{\url{https://aura-tta.onrender.com/}}

\section{Method}
\label{sec:method}

We divide our method in two parts: building a decoder to reconstruct audio from SSL representations, and effectively training a text-conditioned generative model in SSL representation space. We describe both below.

\subsection{Audio reconstruction from SSL representations}\label{sec:aura-dec}

To build AuRAs, we pair two pretrained audio SSL encoders, BEATs~\cite{beats} and MATPAC++~\cite{matpac++}, with decoders trained to reconstruct the input waveform from their frozen representations. Both MATPAC++ and BEATs use ViT~\cite{vit} backbones operating on log-Mel spectrograms. They are pretrained on general audio using AudioSet~\cite{audioset}, and achieve strong performance on general audio classification~\cite{matpac++}, which makes them suitable candidates for studying the RAE hypothesis in audio.

Like other SSL audio encoders, BEATs and MATPAC++ are trained on log-Mel spectrograms of audio signals sampled at $16\,$kHz. This creates two challenges for AuRAs. First, simple inversion of representations with a plain ViT decoder as done in image RAEs would map embeddings to log-Mel spectrograms, and therefore require an additional vocoder module to reconstruct the waveforms. Since modern neural vocoders are tied to the hop size and number of Mel bands they were trained on, using them would require matching front-ends between the selected vocoder and SSL encoder. Additionally, it would also impose an external performance ceiling (the vocoder's) on the decoder. Second, the $16\,$kHz sample rate limits the SSL representations to information below $8\,$kHz, meaning that generating signals at sample rates higher than $16\,$kHz would require solving an additional bandwidth extension problem. 

We provide a solution to the first problem using diffusion transformer (DiT) decoders. As shown in Figure~\ref{fig:reconstruction}, in our approach, we encode the audio signal $\mathbf{x}$ with an SSL encoder, and then train a DiT conditioned on the SSL embedding $\mathbf{z}$ to obtain the complex short-time Fourier transforms (STFTs) $\mathbf{X}$ of the original signal. We employ a LightningDiT~\cite{lightning-dit} backbone for the decoders, using cross-attention over SSL embeddings for conditioning. Complex STFTs are compressed with the magnitude compression scheme of~\cite{music2latent} and treated as two-channel \mbox{(real\,/\,imag.)} images. The decoders are trained using a standard rectified flow~\cite{rectified-flow} objective $\mathcal{L}^d_{\mathrm{FM}} = \mathbb{E}_{\mathbf{x}, t_d}[|| \hat{\mathbf{v}}_d(\mathbf{X}_{t_d}, t_d, \mathbf{z}) - (\varepsilon_d - \mathbf{X})||^2]$, where $\hat{\mathbf{v}}_d(\mathbf{X}_{t_d}, t_d, \mathbf{z})$ is the flow velocity estimated by the neural network at timestep $t_d$ and $\varepsilon_d$ is a $\mathcal{N}(0, \mathbf{I})$ noise sample. Following~\cite{music2latent}, a frequency gain dependent on flow timestep $t_d$ is applied to the complex STFTs, since magnitudes at low frequencies tend to be higher than those at high frequencies at low noise levels. The inverse STFT is used to obtain the final waveforms.

Our focus in this work is on the ability of the generative decoder to reconstruct audio from the proposed conditioning, rather than on high-bandwidth waveform synthesis itself. We therefore operate at $16\,$kHz, which preserves the acoustic content already in the SSL embeddings while keeping the generation problem and computational cost tractable. Extending the decoder to higher sample rates, such as $44.1\,$kHz, would additionally require generating high-frequency detail and is largely orthogonal to the conditioning mechanism in our decoders. We leave this extension to high-fidelity TTA generation for future work.

\subsection{TTA in SSL representation spaces}\label{sec:aura-gen}

Generation with AuRAs is done as in LDMs, simply replacing autoencoder latents with SSL representations. As shown in Figure~\ref{fig:generation}, to generate new audio samples, we train a DiT to produce new SSL embeddings conditioned on a text caption. The generated embedding $\hat{\mathbf{z}}$ is then mapped to an audio signal $\hat{\mathbf{x}}$ using our frozen decoder and the inverse STFT. We also employ a LightningDiT for our generative backbone, and use cross-attention on the T5~\cite{t5}\footnote{\url{https://hf.co/google-t5/t5-base}} embedding $\mathbf{c}$ of the caption for conditioning. Classifier-free guidance (CFG)~\cite{cfg} is used for better caption adherence. Similarly to the decoder, the generator is trained with a rectified flow objective \mbox{$\mathcal{L}^g_{\mathrm{FM}} = \mathbb{E}_{t_g}[|| \hat{\mathbf{v}}_g(\mathbf{z}_{t_g}, t_g, \mathbf{c}) - (\varepsilon_g - \mathbf{z})||^2]$}, with $\hat{\mathbf{v}}_g(\mathbf{z}_{t_g}, t_g, \mathbf{c})$ the estimated velocity field, $t_g$ the generator timestep, and $\varepsilon_g$ a Gaussian noise sample.

In the SSL representation spaces, targets are no longer two-dimensional, but rather flattened 1D sequences of continuous tokens. BEATs and MATPAC++ encoders produce sequences of $N_T$ groups of $N_F$ $D$-dimensional tokens, where $N_T$ and $N_F$ are the number of time and frequency patches in the log-Mel spectrogram. Since~\cite{rae} shows the transformer width $W$ to be a hard constraint for generation in RAE spaces, we set $W \geq D$. Unlike~\cite{rae}, we maintain width across all transformer blocks and omit the proposed decoupled diffusion transformer (DDT)~\cite{ddt} head, which increases the model's effective width without widening the whole backbone. Because we use the same generator architecture across generation spaces in our experiments (Section~\ref{sec:experiments}), adopting it could be problematic. In~\cite{rae}, the DDT head benefit is shown to grow with encoder dimensionality, so a shared DDT head could plausibly advantage the high-dimensional SSL spaces over the lower-dimensional baseline latents.

\section{Experiments}
\label{sec:experiments}

To validate AuRAs, we train generative backbones as described in Section~\ref{sec:aura-gen} on representations from MATPAC++, BEATs, and three autoencoder baselines. The resulting TTA pipelines are evaluated for overall output quality, caption adherence, and robustness. We also evaluate the reconstruction quality of AuRAs versus the baselines.

\vspace{0.10cm}
\noindent\textbf{Datasets.} The AuRA decoders are trained on the full AudioSet dataset, excluding samples belonging to the AudioCaps~\cite{audiocaps} dataset. For the generative models, we use a combination of WavCaps~\cite{wavcaps} and the train and validation splits of AudioCaps. The test split of AudioCaps is kept as a separate test set. MATPAC++ and BEATs encoders are trained on the AudioSet train split, which has a small overlap with the AudioCaps test set. Although AudioCaps-test is isolated from decoder and generator training, to rule out any possibility of leakage, we also separate $847$ caption-audio pairs from the Freesound, BBC Sound Effects, and SoundBible parts of WavCaps as an additional test set (WavCaps-test). For the validation of the generative model and the selection of CFG scaling parameters, we use a set of $1024$ randomly selected caption--audio pairs from the WavCaps + AudioCaps combination. AudioSet and AudioCaps are sampled at $16\,$kHz, while WavCaps is sampled at $32\,$kHz.

\vspace{0.10cm}
\noindent\textbf{Baselines.} We choose the Stable Audio VAE (SA-VAE)~\cite{stable-audio}, SAME~\cite{same}, and the AudioLDM2 VAE (A-VAE)~\cite{audioldm2} as baselines. The SA-VAE is used in the TTA pipelines of~\cite{stable-audio, stable-audio-2, tango-flux}, and is one of the most popular autoencoders for audio LDMs. SAME is the successor to the SA-VAE and incorporates additional losses to account for semantic structure in the latent space; it has been successfully used for sound effects generation in~\cite{stable-audio-3}. Like the BEATs and MATPAC++ encoders, the A-VAE encodes log-Mel spectrograms of $16\,$kHz signals, which makes it a natural point of comparison; it is paired with a pretrained HiFi-GAN~\cite{hifi-gan} vocoder to recover waveforms.

\vspace{0.10cm}
\noindent\textbf{Experimental setup.} We use the pretrained checkpoints of all baselines, taking the small version of SAME. For BEATs and MATPAC++, we pick the \textit{iter3} and AudioSet pretrain checkpoints, respectively. BEATs, MATPAC++, and the A-VAE operate at $16\,$kHz, while SAME and the SA-VAE were trained at $44.1\,$kHz. To train the generator, we resample encoder inputs to each model's native rate, without bandwidth extension, preserving the operating conditions of the pretrained autoencoders. All outputs are low-passed at $8\,$kHz before evaluation, so every pipeline is compared over the same frequency band, while allowing the higher-rate models to operate on their native full-band inputs.

\vspace{0.10cm}
\noindent\textbf{AuRA decoder details.}
AuRA decoders are built using $12$ transformer layers with hidden size $W = 1024$, MLPs of dimension $4096$ and SwiGLU activation functions, query-key normalization, and RMSNorm. For positional encoding (PE), a combination of additive 2D sinusoidal PE and rotary positional encoding (RoPE) is used. Timesteps are sampled during training using a log-Normal distribution of mean zero and unit standard deviation, and injected into the model using $1024$-d Gaussian Fourier embeddings and AdaLN-zero. The timestep-dependent frequency gain is calculated through linear projection of the timestep embeddings. We use $50$ Euler steps at inference time. The decoders are trained on STFTs computed with Hann windows, 1024 FFT points, and window length and hop size of 1024 and 256 samples. The Nyquist bin is dropped, and the segment length is set to $9.968\,$s so that the resulting $512 \times 624$ \mbox{(freq.\ $\times$ time)} STFTs divide evenly into $16 \times 8$ patches. We apply the magnitude compression scheme of~\cite{music2latent} with $\alpha = 0.65$ and $\beta = 0.34$. With this segment length, BEATs and MATPAC++ embeddings are sequences of $496$ and $315$ $768$-d tokens, respectively. The decoders are trained for $144000$ iterations with batch size $256$, using the AdamW optimizer with a constant learning rate of $10^{-4}$. Both decoders have around $275$M parameters.

\vspace{0.10cm}
\noindent\textbf{Generator details.}
For the generators, we use the same DiT architecture as the AuRA decoders, replacing the patch projection with a linear projection from $D = 768$ to $W = 1024$, and substituting the 2D sinusoidal PE and RoPE with their 1D counterparts. We train generators on audio segments of the same length as the decoder, resulting in sequences of $214$ $64$-d tokens for the SA-VAE, $108$ $256$-d tokens for SAME, and $250$ $128$-d tokens for the A-VAE. Training timesteps are sampled from the same log-Normal distribution as the decoder and shifted by $s = \sqrt{m /n}$ following~\cite{rae}, with $m$ the effective data dimension (number of tokens times their dimension) of each space and $n = 4096$. We clamp $s$ to $[\mathrm{e}^{0.5}, \mathrm{e}^{1.15}]$, following \cite{stable-audio-3}. As in the decoders, we use $50$ Euler steps at inference time. All generators are trained with a CFG dropout probability of $0.1$. For each space, we select the CFG scale from $\{1.0, 3.0, 5.0, 7.0, 9.0, 11.0\}$ by evaluating on the validation set; the selected scales are reported in Table~\ref{tab:tta-results}. The generators are all trained for $188380$ iterations with batch size $256$, using the AdamW optimizer with a constant learning rate of $10^{-4}$. All generators have approximately $275$M parameters, with a maximum difference of $1.4$M due to the representation-specific input and output projections.

\vspace{0.10cm}
\noindent\textbf{Reconstruction metrics.} We report the MR-STFT loss~\cite{parallel-wave-gan} (signal-based) and ViSQOL~\cite{visqol} (perceptual) as reconstruction quality metrics. The MR-STFT loss is calculated using $\{512, 1024, 2048\}$ FFT points and window lengths and hop sizes of $\{350, 600, 1250\}$ and $\{70, 120, 250\}$ samples, respectively.

\vspace{0.10cm}
\noindent\textbf{Standard generation metrics.} The overall generation quality is evaluated with FAD~\cite{fad} on LAION-CLAP~\cite{laion-clap} embeddings, while caption adherence is evaluated using CLAP scores, i.e., the cosine similarity between CLAP embeddings of captions and generated audio. Following~\cite{stable-audio-3, tango-flux}, we use the \textit{htsat-unfused} CLAP checkpoint.\footnote{\url{https://hf.co/laion/clap-htsat-unfused}}

\vspace{0.10cm}
\noindent\textbf{Additional evaluation.} For additional evaluation of AuRAs against the baselines, we also report: (i) CLAP scores on the $100$ test set captions with the largest average T5 embedding cosine distance to their $10$ nearest neighbors in the training set; and (ii) the FAD and CLAP curves per training step and floating point operations (FLOPs). Intuitively, the first evaluates generalization to test captions that are semantically distant from the training captions. We mean-pool T5 embeddings of captions to get a single embedding of train/test set entries before calculating distances. The second evaluates the performance gain per additional training step or FLOP, and is an evaluation of how AuRAs scale. Since SSL encoders produce longer token sequences, each generator forward pass is more expensive in compute for AuRAs, which motivates examining how performance improves per FLOP. Similarly, due to the different nature of the generation spaces, it is expected that convergence rates differ between AuRAs and the baselines, motivating the per-training-step analysis. 

\section{Results}
\label{sec:results}

\begin{table}[t]
\centering
\footnotesize
\setlength{\tabcolsep}{3pt}
\begin{tabular}{@{}lcccc@{}}
\toprule
& \multicolumn{2}{c}{AudioCaps-test} & \multicolumn{2}{c}{WavCaps-test} \\
\cmidrule(lr){2-3} \cmidrule(lr){4-5}
Model & MR-STFT $\downarrow$ & ViSQOL $\uparrow$ & MR-STFT $\downarrow$ & ViSQOL $\uparrow$ \\
\midrule
SA-VAE      & $\mathbf{0.84 \pm 0.01}$ & \underline{$4.44 \pm 0.01$} & \underline{$0.86 \pm 0.02$} & \underline{$4.44 \pm 0.01$} \\
SAME        & \underline{$0.84 \pm 0.02$} & $4.33 \pm 0.01$ & $\mathbf{0.80 \pm 0.02}$ & $4.36 \pm 0.02$ \\
A-VAE       & $0.96 \pm 0.01$ & $\mathbf{4.55 \pm 0.01}$ & $1.05 \pm 0.08$ & $\mathbf{4.47 \pm 0.02}$ \\
\midrule
BEATs       & $1.51 \pm 0.02$ & $3.87 \pm 0.02$ &  $1.52 \pm 0.03$ & $3.96 \pm 0.02$  \\
MATPAC++    & $1.14 \pm 0.01$ & $4.11 \pm 0.01$ & $1.29 \pm 0.05$ & $4.09 \pm 0.02$  \\
\bottomrule
\end{tabular}
\caption{Reconstruction quality on AudioCaps-test and WavCaps-test, with standard error of the mean (SEM) $95\%$ confidence intervals (CIs). Best in bold, second-best underlined.}
\label{tab:rec-results}
\vspace{-0.3cm}
\end{table}
\noindent\textbf{Reconstruction.} Table~\ref{tab:rec-results} shows the reconstruction results for MATPAC++, BEATs, and the baselines. For the SSL decoders, we decode each representation once using a single random initial noise sample. MATPAC++ and (in particular) BEATs obtain higher MR-STFT losses and lower ViSQOL scores than the baselines. However, as observed in~\cite{rae}, this does not necessarily imply worse generation quality. With image RAEs, the best performing SSL encoder for generation was not the one with the best reconstruction quality. Similarly, our generation results below show that AuRAs with BEATs and MATPAC++ can achieve competitive, if not superior, TTA generation quality despite reconstruction performance. 

\vspace{0.10cm}
\noindent\textbf{Generation.} Table~\ref{tab:tta-results} shows the TTA results for AuRAs and the baselines. The MATPAC++ AuRA obtains the best FAD scores on both datasets, followed by BEATs. As a distributional metric, FAD is an indicator of overall audio quality; AuRA advantage in it on both datasets shows that SSL representations can obtain better general TTA quality than conventional autoencoders. Concerning CLAP scores, which measure caption adherence, the BEATs AuRA achieves the best results on WavCaps, followed closely by MATPAC++. On AudioCaps, A-VAE leads, but BEATs results are competitive, with a significant overlap in confidence intervals. Under a common generative backbone, AuRAs therefore achieve performance comparable or superior to the baseline autoencoders, despite the difference in reconstruction quality. This highlights the finding that reconstruction quality alone is not a good proxy for how well a space supports audio generation. We hypothesize that semantic modeling in SSL representations can significantly impact generation and believe that such spaces should be further explored for generative tasks in audio. 
\begin{table}[htb]
\centering
\footnotesize
\setlength{\tabcolsep}{4pt}
\begin{tabular}{@{}lcccc@{}}
\toprule
& \multicolumn{2}{c}{AudioCaps-test} & \multicolumn{2}{c}{WavCaps-test} \\
\cmidrule(lr){2-3} \cmidrule(lr){4-5}
Model (CFG) & FAD $\downarrow$ & CLAP $\uparrow$ & FAD $\downarrow$ & CLAP $\uparrow$ \\
\midrule
SA-VAE (5.0)    & 0.070 & 0.528$\pm$0.004 & 0.095 & 0.474$\pm$0.010 \\
SAME (9.0)      & 0.149 & 0.496$\pm$0.004 & 0.177 & 0.415$\pm$0.010 \\
A-VAE (5.0)     & 0.053 & \textbf{0.557$\pm$0.004} & 0.112 & 0.482$\pm$0.010 \\
\midrule
BEATs (5.0)     & \underline{0.048} & \underline{0.550$\pm$0.004} & \underline{0.083} & \textbf{0.494$\pm$0.010} \\
MATPAC++ (5.0)  & \textbf{0.043} & 0.538$\pm$0.004 & \textbf{0.072} & \underline{0.493$\pm$0.010} \\
\bottomrule
\end{tabular}
\caption{TTA results at $50$ sampling steps. CLAP scores are shown with SEM $95\%$ CIs; FAD values are set-level point estimates. Best in bold, second-best underlined.}
\label{tab:tta-results}
\end{table}

\begin{figure}[htb]
  \centering
  \includegraphics[width=\columnwidth]{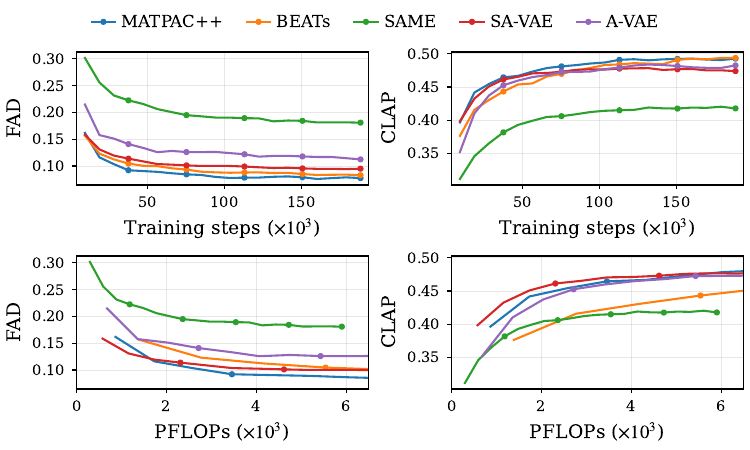}
  \caption{FAD and CLAP as a function of training steps and compute.}
  \label{fig:convergence}
\end{figure}

\noindent\textbf{Scaling.} Fig.~\ref{fig:convergence} shows FAD and CLAP scores against training epochs (top) and training forward pass FLOPs (bottom) on WavCaps. Since batch size is fixed for all TTA pipelines, the former compares TTA pipelines under an equal-data budget while the latter compares them under an equal-compute budget. On equal-compute, TTA with MATPAC++ sits below the baselines on FAD, and is competitive on CLAP scores; the BEATs AuRA is competitive on FAD, but not as much on CLAP due to longer token sequences. On equal-data both AuRAs are comparable or better than baseline autoencoders, with MATPAC++ slightly superior on both FAD and CLAP. Overall, AuRAs can provide faster convergence per training sample, but longer token sequences in SSL representations and their compute costs need to be taken into account. 

\begin{table}[htb]
\centering
\footnotesize
\setlength{\tabcolsep}{6pt}
\begin{tabular}{@{}lcc@{}}
\toprule
Model (CFG) & AudioCaps-test & WavCaps-test \\
\midrule
SA-VAE (5.0)    & 0.444$\pm$0.033 & \underline{0.451$\pm$0.026} \\
SAME (9.0)      & 0.418$\pm$0.032 & 0.404$\pm$0.030 \\
A-VAE (5.0)     & \underline{0.490$\pm$0.032} & 0.438$\pm$0.031 \\
\midrule
BEATs (5.0)     & \textbf{0.518$\pm$0.025} & \textbf{0.476$\pm$0.030} \\
MATPAC++ (5.0)  & 0.475$\pm$0.030 & 0.447$\pm$0.030 \\
\bottomrule
\end{tabular}
\caption{CLAP$_{100}$ ($\uparrow$), the CLAP score on the $100$ most underrepresented samples of each test set. SEM $95\%$ CIs. Best in bold, second-best underlined.}
\label{tab:clap100}
\vspace{-0.10cm}
\end{table}
\vspace{0.10cm}
\noindent\textbf{Robustness.} We additionally calculate CLAP$_{100}$ (Table~\ref{tab:clap100}), the average CLAP score on the $100$ samples of AudioCaps-test and WavCaps-test with the largest average T5 embedding cosine distance to their 10 nearest neighbors in the training set. This serves as a proxy robustness metric for uncommon captions. TTA with BEATs achieves the best CLAP$_{100}$ in both test sets, while MATPAC++ is competitive with second-best results from autoencoder baselines. Furthermore, BEATs and MATPAC++ have the smallest performance drop from their original CLAP scores from Table~\ref{tab:tta-results}, showcasing AuRA robustness in comparison with the baselines. Robustness to outlier captions is further indication that semantic content in SSL representations could benefit generation in these spaces.

\section{Conclusion}
\label{sec:conclusion}

This paper presented a study on AuRAs and their applicability to TTA. We trained diffusion decoders to convert representations from two SSL encoders back to the original signals, and evaluated AuRAs as a drop-in replacement for conventional LDM autoencoders in TTA using a common text-conditioned generative backbone. We show that AuRAs are a viable alternative to autoencoders used in traditional audio LDMs. Due to longer token sequences generated by SSL encoders, AuRAs impose a higher compute cost per forward pass, but scaling experiments show competitive performance. More interestingly, AuRAs show robustness to outlier text captions, and exhibit similar or superior generation quality to existing autoencoders when compared using the same generative backbone, despite inferior reconstruction capabilities. We suppose that the relative decoupling of reconstruction and generation quality could come from semantic properties of the SSL representations and see this as a promising research direction for improving audio generative models. 

% \section{Acknowledgements}
% This work was performed using HPC resources from GENCI-IDRIS (Grant 20XX-AD011016124R1). Claude AI was used for language editing, figure drafts, and to accelerate code development.

% Bibliography
% -------------------------------------------------------------------------
\bibliographystyle{IEEEbib}
\bibliography{refs}

\end{document}